\documentclass{article}
\usepackage{graphicx}

\title
{Analytical description of the coherent structures within the
hyperbolic generalization of Burgers equation}
\author{ V.A. Vladimirov
and E.V. Kutafina \\ University of Science and Technology\\
Faculty of Applied Mathematics \\ Al Mickiewicza 30, 30-059 Krakow, Poland \\ E-mail vsan@rambler.ru \\[2ex] }
\begin{document}
\newtheorem{thm}{Theorem}
\newtheorem{lm}{Lemma}
\maketitle
\begin {abstract}

 We present new periodic,  kink-like and  soliton-like travelling wave
solutions to the hyperbolic generalization of Burgers equation. To
obtain them, we employ the classical and generalized symmetry
methods and the ansatz-based approach
\end{abstract}

\section{ Introduction }

Actually there are no analytical methods enabling to  solve  an
arbitrary initial or boundary value problems for  nonlinear PDEs
that are not completely integrable. The majority of evolutionary
PDEs used to simulate non-linear transport phenomena is not
completely integrable. Yet it is of common knowledge that under
certain conditions coherent structures formation take place in
open dissipative systems, being simulated by PDEs, that are not
completely integrable.  The analytical description of the coherent
structures within the non-integrable models is of great interest,
since  analytical form of  solution is more preferable in
analyzing structures formation and evolution and, besides, they
are widely used as a starting point for various asymptotic
methods, for testing the numerical schemes and facilitating the
stability analysis.

To obtain exact solutions to non-linear PDEs, various methods are
used. Besides the classical symmetry reduction  scheme (see e.g.
\cite{ovs, olver1}), which is not effective in obtaining solutions
with the given properties, there are employed  techniques based on
choosing a proper transformation (or ansatz), turning the problem
of finding out exact solutions to the algebraic one
\cite{fan,nikbar,baryur}. In our previous study \cite{vladku04} we
employed a certain dialect of the the direct algebraic method to
the hyperbolic generalization of Burgers equation (GBE). Yet in
the following publications \cite{vladku05,vskup} we showed that
within this methodology it is impossible to obtain the the
solitary wave solution of GBE, occurring to exist for certain
values of the parameters \cite{vladku05}. This served us as a
motivation for the developing the effective methods of
approximations \cite{vladku05,vskup} of solutions describing the
coherent structures. Actually we return to the subject of the
analytical description of the coherent structures within the GBE
model, using for this purpose quite different technique, namely, a
generalized symmetry and the combination of the classical symmetry
reduction with an ansatz-based method \cite{clarkson}.

The structure of the study is following. In section 2 we show how
it is possible, combining the classical reduction with the
Hirota-like ansatz, to linearize equation describing a set of
travelling wave solution to GBE and  obtain on this basis new
periodic, kink-like and soliton-like solution.  In the following
section we show how it is possible to gain the linearization of
the reduced GBE, using the Painlev\'{e} test. In section 4, using
the idea of conditional symmetry, we obtain a wide class of exact
solutions to GBE, containing the functional arbitrariness. In
section 5 we give some examples of such solutions. In the final
part of the study we make some conclusions and general remarks.

\section{Solutions to GBE obtained via the classical symmetry reduction and Hirota-like
ansatz}\label{sec:clasandhir}

Let us consider a hyperbolic generalization of Burgers equation
\cite{makar1, vladku04}:
%\end{center}
\begin{equation}\label{genb1} \tau u_{tt}+Auu_x+Bu_t+H\,u_x-\kappa
u_{xx}=f(u)=\lambda\,(u-m_1)(u-m_2)(u-m_3). \end{equation}

\noindent We are going to look for the wave patterns among the set
of travelling wave (TW) solutions, i.e.  solutions invariant with
respect to translations generators $\frac{\partial}{\partial\,t}$
and $\frac{\partial}{\partial\,x}$. So it is quite natural to
employ the classical reduction scheme \cite{ovs, olver1}, and
consider an over-determined system containing action of the
generator $\hat
X=\frac{\partial}{\partial\,t}\,-\,\mu\,\frac{\partial}{\partial\,x}$
on the solutions of GBE:
\begin{equation}\label{GBE}\tau u_{tt}+Auu_x+Bu_t+H\,u_x-\kappa u_{xx}=f(u),
\end{equation}
\begin{equation}\label{classym}
\hat X\,\left[u(t,\,x)-u\right]=u_t-\mu\,u_x={0}.
\end{equation}
The classical theorems \cite{ovs, olver1} assure that system
(\ref{GBE})--(\ref{classym}) is compatible and has non-trivial
solutions.
In fact, solving equation (\ref{classym}) we obtain an ansatz
\begin{equation}\label{anstw}
u(t,\,x)=U(\xi), \qquad \xi=x+\mu\,t.
\end{equation}
Inserting ansatz (\ref{anstw}) into (\ref{GBE}), we  obtain  the
following ODE:
\begin{equation}\label{redode}
(\tau\,\mu^2-\kappa)\,U^{''}(\xi)+(H+B\,\mu)\,U'(\xi)+
A\,U(\xi)\,U'(\xi)=\lambda\,(U(\xi)-m_1)(U(\xi)-m_2)(U(\xi)-m_3).
\end{equation}
The problem is that, generally speaking, equation (\ref{redode})
is non-integrable, so, employing the classical reduction scheme we
are not able to obtain analytical description to wave patterns.

In the following  we use this scheme together with the Hirota-like
ansatz
\begin{equation}\label{hirotalike} U(\xi)=\frac{\Psi'(\xi)}{\Psi(\xi)},\end{equation}
and show that  under certain conditions such combination leads to
a linear ODE. Inserting (\ref{hirotalike}) into (\ref{redode}), we
obtain the equation
\begin{equation}\label{eq:forphi}\begin{array}{ll} \Psi(\xi)^2\bigl[\lambda\,m_1m_2m_3\Psi(\xi)-
\lambda\,(m_2m_3+m_1m_2+m_1m_3)\Psi'(\xi)+(H+B\mu)\Psi''(\xi)+\\\\
(\mu^2\tau-\kappa)\Psi'''(\xi)\bigr]
+\Psi(\xi)\,\Psi'(\xi)\bigl[\Psi''(\xi)\left(
A+3\kappa-3\mu^2\tau\right)
-\Psi'(\xi)\left(H- \right.\\\\
\left.-\lambda\,(m_1+m_2+m_3)+B\mu\right)\bigr]
+\left[\Psi'(\xi)\right]^3\left(A-\lambda\,-2\kappa+2\mu^2\tau\right)=0.\end{array}
\end{equation}
%\end{document}
One easily gets convinced by the direct inspection that the
following statement holds:
\begin{lm}
\label{lem1}
 Ansatz (\ref {hirotalike}) leads to the
linear ODE
\begin{equation}\label{a}\lambda\,\left[m_1m_2m_3\Psi-(m_2m_3+m_1m_2+m_1m_3)\Psi'\right]+
(H+B\mu)\Psi''+(\mu^2\tau-\kappa)\Psi'''=0,\quad
\end{equation} provided that following conditions are fulfilled:
\begin{equation}\label{b} A+3\kappa-3\mu^2\tau=0, \end{equation}
\begin{equation}\label{c}H-\lambda\,(m_1+m_2+m_3)+B\mu=0, \end{equation}
\begin{equation}\label{d}A-\lambda-2\kappa+2\mu^2\tau=0.
 \end{equation}
\end{lm}

Equation (\ref{a}) is a third-order ordinary linear differential
equation.  Its solutions depend on the roots $\left\{\sigma_k
\right\}_{k=1,\,2,\,3}$ of corresponding characteristic equation.
In case when the conditions (\ref{b})--(\ref{d}), are fulfilled,
the characteristic equation is as follows:
\begin{equation}\label{chareq}
\sigma^3-\left( {m_1+m_2+m_3}\right)\,\sigma^2+\left(
m_2\,m_3+m_1\,m_2+m_1\,m_3\right)\,\sigma-m_1\,m_2\,m_3=0.
\end{equation} The roots of equation (\ref{chareq}) coincide
with the numbers $m_1$, $m_2$ and $m_3$, being the  roots of
equation $f(u)=0$. Below we consider four distinct cases.

\begin{figure}
\includegraphics[width=3 in, height=1.5 in]{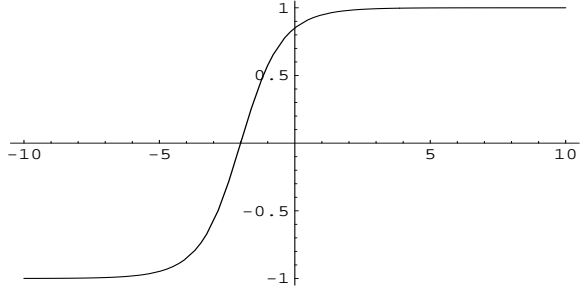}
\caption{An example of TW solution described by the formula
(\ref{threeroots})}\label{bia1}
\end{figure}

{\bf Case I.} For  $m_1\neq m_2\neq m_3 \neq m_1$ the general
solution of (\ref{a}) takes on the form
%\end{center}
$$\Psi(\xi)=e^{m_1\,\xi}c_1+e^{m_2\,\xi}c_2+e^{m_3\,\xi}c_3.$$
In the following  we put  $c_1=1$. With this assumption we
obtain the  expression
\begin{equation}\label{threeroots} u(t,\,x)=\frac{m_1e^{m_1\xi}+c_2m_2e^{m_2\xi}+
c_3m_3e^{m_3\xi}}{e^{m_1\xi}+c_2e^{m_2\xi}+c_3e^{m_3\xi}},\quad
\xi=x+\mu t.\end{equation} For positive $c_2\,\, \mbox{and}\,\,
c_3$  formula (\ref{threeroots}) describes a kink-like solution
(see fig~\ref{bia1}).

{\bf Case II:} For $m_1\neq m_2=m_3$
\[
\Psi(\xi)=e^{m_1\xi}+e^{m_2\xi}(c_2+c_3\xi)
\]
and using the formula (\ref{hirotalike}) we get the solution
\begin{equation}\label{tworoots} u(t,\,x)=\frac{m_1e^{m_1\xi}+c_3e^{m_2\xi}(c_2\,m_2+c_3+m_2\,c_3\xi)}
{e^{m_1\xi}+e^{m_2\xi}(c_2+c_3\xi)}.\end{equation}

\begin{figure}
\includegraphics[width=0.3\textwidth]{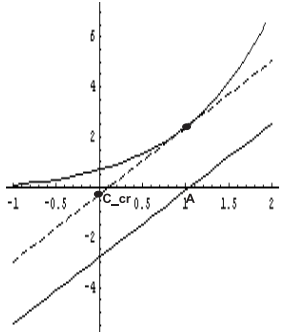}
\\
\caption{Case II:   nonsingular solutions. Solid lines correspond
to the plots of functions $f_1(\xi)=e^{(m_1-m_2)\xi}$ and
$f_2(\xi)=-(c_2+c_3\xi)$. Dashed line is a tangent to $f_1(\xi),$
being parallel to $f_2(\xi)$ }\label{fig:2}
\end{figure}

Asymptotic analysis  shows, that  solution (\ref{tworoots}) tends
to either $m_1$ or to $m_2$  as  $\xi\rightarrow\pm\,\infty$. The
condition of non-singularity of the solution reads as follows:
$\forall \xi \in R:$ $e^{m_1\xi}+e^{m_2\xi}(c_2+c_3\xi)>0$ or, in
other words,
\begin{equation}\label{cond1} e^{(m_1-m_2)\xi}>-(c_2+c_3\xi).\end{equation} It is possible when
$m_1-m_2$  is positive, and  $c_3$ is negative or when $m_2-m_1$,
and $c_3$ are positive. In fig.~\ref{fig:2}  the first possibility
is  illustrated. Note that the second one is symmetric. It is
evident from the analysis of fig.~\ref{fig:2} that condition
(\ref{cond1}) is satisfied in case when $-c_2<c_{cr}$. One can
easily verify that $c_{cr}$ is expressed by the formula
\[
\qquad c_{cr}=-A\left[c_3+e^{m_1-m_2}\right],\]
providing that $A=\frac{1}{m_1-m_2}\,ln\,\frac{c_3}{m_2-m_1}.$

\begin{figure}
\includegraphics[width=3 in, height=1.5 in]{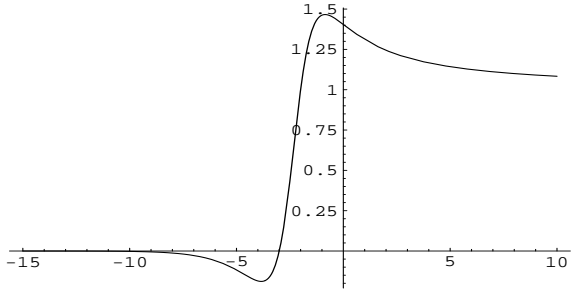}
\caption{An example of TW solution described by the formula
(\ref{tworoots})}\label{bib}
\end{figure}

\begin{figure}\includegraphics[width=3 in, height=1.5 in]{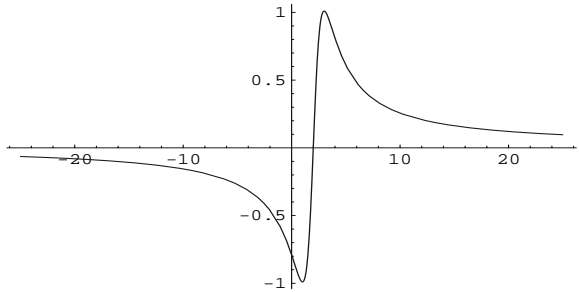}
\caption{An example of TW solution described by the formula
(\ref{heavytail})}\label{bic1A} \end{figure}

{\bf Case III:} For  $m_1=m_2=m_3=m$
\[
\Psi(\xi)=e^{m\xi}\,\left[c_3+\xi\,(c_2+\xi)\right]
\]
and using (\ref{hirotalike}) we get the solution
 \begin{equation}\label{heavytail}
u(t,\,x)=m+\frac{c_2+2\xi}{c_3+\xi(c_2+\xi)},\end{equation}
Formula (\ref{heavytail}) describes a a solitary wave with "heavy"
tails (fig.~\ref{bic1A}), providing that inequality
$c_2^2-4\,c_3<0$ holds.

\begin{figure}
\includegraphics[width=3 in, height=1.5 in]{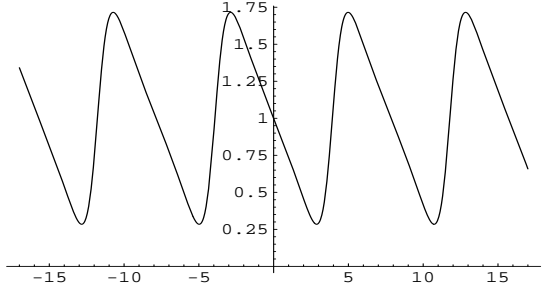}
\caption{Periodic TW solution to (\ref{GBE}) described by the
formula (\ref{complexroots})}\label{bid1}
\end{figure}

{\bf Case IV:} For $m_1,\,\,m_2=\alpha \pm\,\beta \,i$,   and real
$m_3,$ we get the solution
\begin{equation}\label{complexroots} u(\xi)=\frac{c_3m_3e^{m_3\xi}+2e^{\alpha \xi}\left[\alpha\,
\mbox{cos}(\beta\xi)-\beta\, \mbox{sin}(\beta\xi)\right]}
{c_3e^{m_3\xi}+2\,e^{\alpha
\xi}\mbox{cos}(\beta\xi)}\end{equation}
Let us notice, that the only possibility for the solution
(\ref{complexroots}) to be non-singular is simultaneous
fulfillment of the  conditions $m_3=\alpha$ and  $|c_3|>2.$ In
this case we get a periodic solution (fig.~\ref{bid1}).

\section{Painlev\'{e} analysis of GBE}\label{sec:painl}

Now we are going to show how the  conditions on constants similar
to (\ref{b})--(\ref{d})  can be obtained using the Painlev\'{e}
test \cite{tabor}. Since we concentrate here merely upon the set
of TW solutions, let's start from the equation (\ref{redode}).

 In the first step we take $U(\xi)=U(x+\mu t)=a[0]\xi^p$ and
 put it into the leading terms of (\ref{redode}). This way we obtain the relation:
\begin{equation}(p-p^2)(\mu^2\tau-\kappa)+A\,p\,\xi^{p+1} a[0]+\lambda\,\xi^{2p+2}a[0]^2=0.\end{equation}
In order that the equation be balanced, we must put
$2(p+1)=p+1$, so $p=-1$.

Next we take $U(\xi)=\sum_{i=0}^{\infty}a[i]\xi^{i-1}$. Equating to
zero coefficients of each power of $\xi$, we get the recurrence
for $a[i]$. The first two expressions are as follows:
\begin{equation}{\kappa\lambda}\,a[0]={A\pm\sqrt{A^2-8\lambda\,(\mu^2\tau-\kappa)}} \end{equation}
\begin{equation}(A-3\lambda\,a[0])a[1]=-(H+B\mu+\lambda\,(m_1+m_2+m_3)a[0]) \end{equation}
Since in  the Painlev\'{e} test being applied to a second order
equation  one of $a[i]$ should be "free", then we can put
\begin{equation}\begin{array}{ll}\label{pain1}
A-3\lambda\,a[0]=0\\
H+B\mu+\lambda\,(m_1+m_2+m_3)a[0]=0.\\
\end{array}
\end{equation}
Solving (\ref{pain1}) we obtain:
\[
\lambda\,=\frac{s^2}{\mu^2\tau-\kappa}, \qquad A=-3s, \qquad
s=\frac{H+B\mu}{m_1+m_2+m_3}.
\]
If we additionally put $a[0]=1$ then we get precisely the
conditions  for which equation (\ref{eq:forphi}) linearizes.

\section{Exact solutions of GBE associated with the conditional
symmetry}\label{sec:condsol}

Now let us consider an over-determined system \begin{eqnarray}
\tau u_{tt}-\kappa
u_{xx}+u\,u_x+u_t=f(u), \label{GBE4} \\
  \hat X\,[u]=\nu\,u_t+u_x\,=\,\Phi(u), \label{condgen1}
  \label{condsym}\end{eqnarray}
where \begin{equation} \hat
X=\nu\partial_t+\partial_x+\Phi(u)\partial_u
\end{equation}
 is a generalized
(conditional) symmetry generator containing an unknown function
$\Phi(u)$.

In accordance with  \cite{WF1,olver3,olver2}, system
(\ref{GBE4}--\ref{condgen1}) will be compatible if it admits the
second prolongation of the operator $\hat X$:
\begin{equation}\label{prolong2}
\hat X_{2}=\hat X+\zeta_{xx}\,\frac{\partial }{\partial
u_{xx}}\,+\zeta_{t}\,\frac{\partial }{\partial
u_t}+\zeta_x\,\frac{\partial }{\partial
u_x}+\zeta_{tt}\,\frac{\partial }{\partial
u_{tt}}+\zeta_{tx}\,\frac{\partial }{\partial u_{tx}}
\end{equation}
where
\[
\zeta_{t}=D_t\,\Phi(u),\quad\zeta_{x}=D_x\,\Phi(u),\quad\zeta_{tt}=D_t\,\zeta_{t},\quad
\zeta_{tx}=D_t\,\zeta_{x},\quad\zeta_{xx}=D_x\,\zeta_{x}
\]
and $D_t,\,D_x $ are total derivatives with respect to
corresponding variables. Applying the operator (\ref{prolong2}) to
equation (\ref{GBE4}), we get:
\begin{equation}\label{defining1}
\zeta_{t}+\tau\,\zeta_{tt}+\Phi(u)\,u_x+u\,\zeta_{x}-\kappa\,\zeta_{xx}-\dot{f}(u)\,\Phi(u)|_{E_i=0}=0,
\end{equation}
where $E_i=0$  means that equation is considered on the manifold,
i.e. when equations (\ref{GBE4})--(\ref{condgen1}) and all
their differential consequences are taken into account.

A passage to the manifold defined by equations (\ref{GBE4})--(\ref{condgen1})
requires some comments. Equation (\ref{condgen1}) and its differential consequences
give us the following conditions:
\begin{eqnarray}
u_x=\Phi(u)-\nu\,u_t, \label{cond}  \\
u_{xx}=\dot{\Phi}(u)\,\Phi(u)-2\,\nu\,\dot{\Phi}(u)\,\,u_t+\nu^2\,u_{tt}.
\label{difcons2}
\end{eqnarray}
Using these formulae and  equation  (\ref{GBE4}), we can also
exclude $u_{tt}$. In fact, inserting
(\ref{cond})--(\ref{difcons2}) into (\ref{GBE4}), we obtain the
equation
\begin{equation}\label{difcons2a}
\left(\tau-\kappa\,\nu^2\right)\,u_{tt}=f(u)+\left(\nu\,u-1-
2\,\kappa\,\nu\,\dot{\Phi}(u)\right)\,u_{t}+\Phi(u)\left(\kappa\,\dot{\Phi}(u)-u\right).
\end{equation}
Here we have two possibilities, depending on whether
$\left(\tau-\kappa\,\nu^2\right)$ is zero or not.

If $\tau-\kappa\,\nu^2\,\neq\,0,$ then, returning to
(\ref{defining1}) and taking account of
(\ref{cond})--(\ref{difcons2}) and (\ref{difcons2a}), we obtain a
first order PDE. Applying the standard procedure of splitting
\cite{WF1,olver3,olver2} and  solving the  system of determining
equations we conclude that $\Phi[u]=0$ and therefore $\hat X$ is a
classical symmetry generator.

Now let us assume that $\tau-\kappa\,\nu^2\,=\,0.$ Using the
formulae (\ref{cond})--(\ref{difcons2}), we can exclude from
(\ref{defining1})  merely the spatial derivatives $u_x$ and
$u_{xx}$. The splitting procedure gives us in this case the
following  system:
\begin{eqnarray}
\kappa\,\left\{{\Phi}(u)\ddot{\Phi}(u)+\left[\dot{\Phi}(u)\right]^2\right\}=
{\Phi}(u)+u\,\dot{\Phi}(u)-\dot{f}(u) \label{pervopr} \\
2\,\kappa\,\nu\,\left\{{\Phi}(u)\ddot{\Phi}(u)+\left[\dot{\Phi}(u)\right]^2\right\}=
\left(\nu\,u-1\right)\dot{\Phi}(u)+\nu\,{\Phi}(u)\label{vtoropr}
\end{eqnarray}
Integrating once equations (\ref{pervopr})--(\ref{vtoropr}), we
get the system
\begin{eqnarray}
\kappa\,\Phi(u)\dot{\Phi}(u)=
u\,{\Phi}(u)-{f}(u)-C_1, \label{calkpervopr} \\
2\,\kappa\,\nu\,{\Phi}(u)\dot{\Phi}(u)=
\left(\nu\,u-1\right)\,{\Phi}(u)+C_2. \label{calkvtoropr}
\end{eqnarray}
Multiplying  (\ref{calkpervopr}) by $2\,\nu$ and extracting the
resulting equation from (\ref{calkvtoropr}), we get the relation
between the functions $\Phi(u)$ and $f(u)$:
\begin{equation}\label{phi}
\Phi\,(u)=\frac{2\,\nu\,\left[f(u)+C_1\right]+C_2}{\nu\,u+1}\,.
\end{equation}

Let us note that function $\Phi\,(u)$  is defined by the Abel-type
equation (\ref{calkvtoropr}). Passing to the new independent
variable $\zeta=\left(\nu\,u-1\right)\,C_2/\left(2\,\kappa\,\nu^2
\right),$ we can bring it to the canonical form \cite{polyzajts}:
\[
{\Phi}(\zeta)\frac{d\,\Phi(\zeta)}{d\,\zeta}=
\alpha\,\zeta\,{\Phi}(\zeta)+1, \qquad
\alpha=\frac{2\,\kappa\,\nu^2}{C_2^2}.
\]
A passage to the new variables
\[
\xi=\Phi-\frac{\alpha}{2}\,\Psi^2\left(\xi \right)
\]
leads to the Riccati equation
\[
\frac{d\,\Psi}{d\,\xi}=\frac{\alpha}{2}\,\Psi^2+\xi.
\]
Solution to this equation is expressed by the following formula
\cite{polyzajts}:
\begin{equation}\label{phibess}
\Psi\left(\xi\right)=\sqrt{\xi}\left[A\,J_{\frac{1}{2q}}\left(\frac{1}{q}\sqrt{\frac{\alpha}{2}}\,\xi^q\,\,
\right)+B\,Y_{\frac{1}{2q}}\left(\frac{1}{q}\sqrt{\frac{\alpha}{2}}\,\xi^q\,
\right) \right], \qquad q=\frac{3}{2},
\end{equation}
where $A,\,\,B$ are constant parameters while $J_m(z),\,\,Y_m(z)$
are the Bessel functions.

So, in accordance with the formulae (\ref{phibess}) and
(\ref{phi}), $f(u)$ is expressed in a complicated way by special
functions. In order to obtain an explicit description to $f(u)$,
we set $C_1=C_2=0$. In this case system
(\ref{calkpervopr})--(\ref{calkvtoropr}) has the following
solution:
\begin{eqnarray}
f(u)=h_0+\left(\nu\,h_0-\frac{1}{4\,\kappa\,\nu^2} \right)\,u-
\frac{1}{8\,\kappa\,\nu}\,u^2+
\frac{1}{8\,\kappa}\,u^3, \label{fpol} \\
\Phi(u)=\frac{\nu\,u^2-2\,u+8\,\nu^2\,h_0\,\kappa}{4\,\nu\,\kappa},\label{phipol}
\end{eqnarray}
where $h_0$ is an arbitrary constant.
%Let us formulate the result obtained as the following statement.
%
Now we are going to formulate the main result of this section.

\begin{thm} \label{Th1}   Suppose that $f(u)$  and $\Phi(u)$
 are given by the formulae (\ref{fpol}) and
(\ref{phipol}) respectively.
%\[
%f(u)=\sum_{k=0}^3\,h_k\,u^k
%\]
%where $\{h_k\}_{k=1}^3$ are given by the formula (\ref{h-koef}).
Then
\begin{itemize} \item system (\ref{GBE4})-(\ref{condsym}) is
compatible; \item  every
solution of (\ref{condsym}) satisfies (\ref{GBE4}).
\end{itemize}
\end{thm}

\section{Examples of conditionally-invariant solutions of
GBE}\label{sec:condeg}

 For
$\nu=\tau=\kappa=1$
%(\ref{condsym})
equation  (\ref{condgen1}) takes the form
\begin{equation}\label{eq:condgensol}
u_t+u_x=\frac{1}{4}\,\left(u^2-2\,u+8\,h_0 \right).
\end{equation}
General solution of this equation depends on whether
$\Delta=1-8\,h_0$ is positive or not. For $h_0<1/8$ solution of
(\ref{eq:condgensol}) is as follows:
\begin{equation}\label{eq:solcondgen1}
u(t,\,x)=\frac{u_2\,G(\omega)\,e^{t\,\frac{\sqrt{\Delta}}{2}}-u_1}{G(\omega)\,e^{t\,\frac{\sqrt{\Delta}}{2}}-1}\,,
\end{equation}
where
\[
u_1=1+\sqrt{\Delta}\,, \qquad u_2=1-\sqrt{\Delta}\,,
\]
$G(\cdot)$ is an arbitrary function of $\omega=x-t$. Putting
$h_0=-1$ and $G(\omega)=-e^{\Gamma(\omega)}$ we obtain the formula
\begin{equation}\label{eq:solcondgen2}
u(t,\,x)=
2\,\frac{2-\mbox{exp}[3\,t/2+\Gamma(\omega)]}{1+\mbox{exp}[3\,t/2+\Gamma(\omega)]}.
\end{equation}
For $h_0=-1$ and $G(\omega)=e^{\Gamma(\omega)}$ we get the
solution
\begin{equation}\label{eq:solcondgen3}
u(t,\,x)=
2\,\frac{\mbox{exp}[3\,t/2+\Gamma(\omega)]+2}{1-\mbox{exp}[3\,t/2+\Gamma(\omega)]}.
\end{equation}

If $h_0>1/8$ then solution to (\ref{eq:condgensol}) is as follows:
\begin{equation}\label{eq:solcondgen4}
u(t,\,x)= 1+\beta\,\mbox{arctg}\left[\frac{\beta\,t}{4}+G(\omega)
\right],
\end{equation}
where $\beta=\sqrt{|1-8\,h_0|}$. This solution is always singular.

If $h_0=1/8$ then solution to (\ref{eq:condgensol}) is as follows:
\begin{equation}\label{eq:h0eqzero}
u(t,\,x)= 1+\frac{1}{G(\omega)-\frac{t}{4}},
\end{equation}
This solution is also singular.

%************************

\begin{figure}
\includegraphics[width=3 in, height=1.5 in]{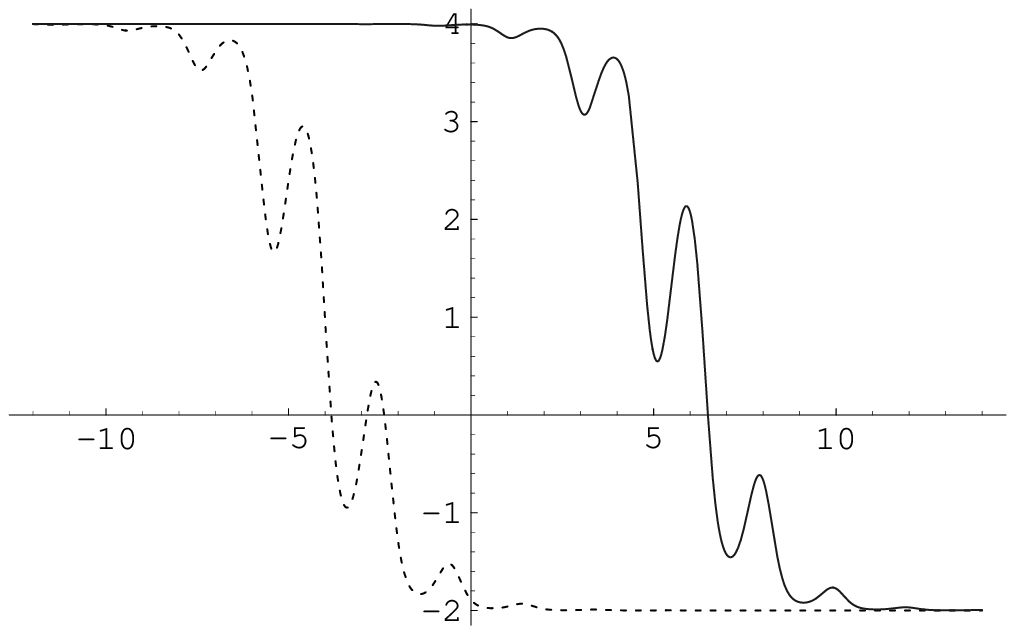}
\caption{Temporal evolution of solution described by formula
(\ref{eq:solcondgen2}) in case when
$\Gamma({\omega})=\mbox{sin}[2.25\, \omega]$ }\label{fig:6}
\end{figure}

\begin{figure}
\includegraphics[width=3 in, height=1.5 in]{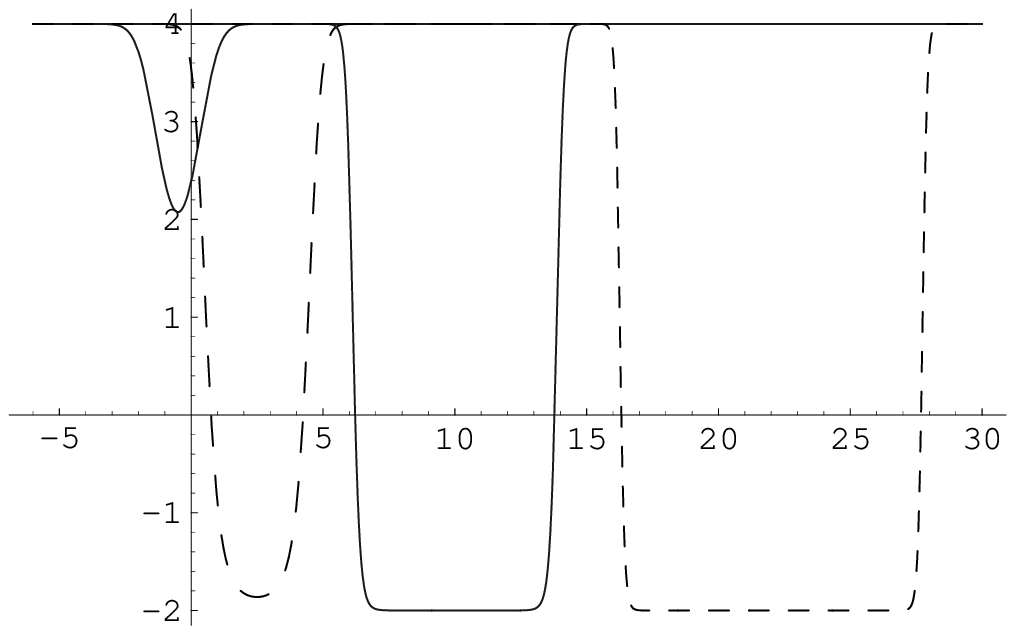}
\caption{Temporal evolution of solution described by formula
(\ref{eq:solcondgen2}) in case when
$\Gamma(\omega)=-\omega^2$}\label {fig:7} \end{figure}

Let us give examples of solutions corresponding to formulae
(\ref{eq:solcondgen2}) and (\ref{eq:solcondgen3}). Thus, inserting
$\quad \Gamma(\omega)=\mbox{sin}\,[2.25\,\omega]$ into equation
(\ref{eq:solcondgen3}), we obtain an oscillating kink-like
solution, shown in fig.~\ref{fig:6}. For
$\Gamma(\omega)=-\omega^2$ this solution produces a "dark" soliton
with a growing support (fig~\ref{fig:7}).

\begin{figure}
\includegraphics[width=3 in, height=1.5 in]{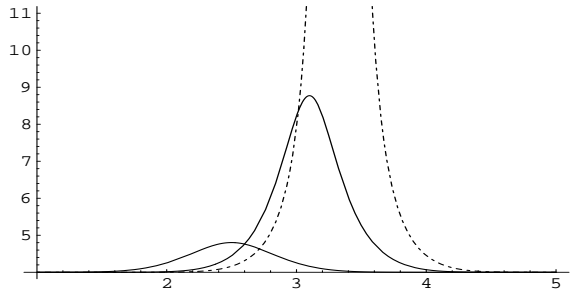}
\caption{A blow-up regime described by the formula
(\ref{eq:solcondgen3})  in case when $\Gamma(\omega)=-3.75
\,\omega^2+5$}\label{fig:8}
\end{figure}

In contrast to (\ref{eq:solcondgen2}), solution
(\ref{eq:solcondgen3}) is always singular. For
$\Gamma(\omega)=-3.75 \,\omega^2+5$ its evolution is shown in
fig.~\ref{fig:8}, in which we see how an initial localized wave
pack grows in amplitude and in a finite time gives rise to a
blow-up regime.

\section{Concluding remarks}

In this study new solutions to GBE, describing wave patterns
formation and evolution have been obtained by means of classical
and generalized symmetry methods and ansatz-based approach. Most
of these solutions cannot be obtained within the dialects of
direct algebraic methods, previously applied to this equation
\cite{vladku04}.

Let us discuss the  solutions obtained in section 2. In the case
when $m_1\neq m_2\neq m_3\neq m_1$ we deal with solutions
described by a rational combination of the exponential functions
which, generally speaking,  do not coincide with the powers of a
single exponential function as it was the case in \cite{vladku04}.
In the following cases considered in section 2, solutions are
described as  rational  combinations of exponential, polynomial
and trigonometric functions and such combinations were never
applied to GBE before.

Solutions obtained in section 4  contain  arbitrary functions,
depending on the linear combination of spatio--temporal variables.
From one hand, it allows, by  a proper choice of these functions,
to  construct a variety of different solutions, such as those
presented in section 5. From the other hand, appearance of an
arbitrary function can be employed to describe a sufficiently wide
family of an initial and (or) boundary value problems. Obtaining
of such solutions has become possible due to employment of
conditional symmetry.

The results presented in section~2 suggest that the possibilities
of obtaining new solutions for GBE within the ansatz-based methods
are not exhausted. It should be noted, however, that effectiveness
of these methods grows in the situation when they are combined
with another methods such as  symmetry reduction or the
Painlev\'{e} test. As it was shown in  section 4, application of
the methods based on conditional symmetry is also very promising.
Let us note in conclusion, that for GBE the most simple
conditional symmetry has been obtained as yet. Our preliminary
investigations show, that there are another conditional symmetry
operators admitted by GBE, which seem to be very promising from
the point of view of obtaining new sets of solutions.

\end{document}